\documentclass[reprint,PRL,twocolumn,showpacs]{revtex4-1}

\usepackage{natbib}
\usepackage{dcolumn}
\usepackage[T1]{fontenc}
\usepackage{amssymb}
\usepackage[fleqn]{amsmath}
\usepackage[latin9]{inputenc}
\usepackage{verbatim}
\usepackage{babel}
\usepackage{pifont}
\usepackage[dvips]{epsfig}
\usepackage{bm}
\usepackage{amsmath}
\usepackage{relsize}
\usepackage[dvips]{graphicx}
\usepackage[usenames,dvipsnames]{color}
\usepackage{float}
\usepackage{setspace}
\usepackage{caption}
\usepackage{subfig}
\usepackage{soul}
\usepackage{ulem}

\usepackage{epstopdf}

\DeclareGraphicsExtensions{.pdf,.jpeg,.png,.eps,.jpg,.gif}

\begin{document}

\title{Antiresonant driven systems for particle manipulation}
\author{M. Florencia Carusela$^{1,2}$}
\email{flor@campus.ungs.edu.ar}
\author{Paolo Malgaretti $^{3,4,5}$}
\email{p.malgaretti@fz-juelich.de}
\author{J. Miguel Rubi $^{6}$}
\email{mrubi@ub.edu}

\affiliation{$^1$ Instituto de Ciencias, Universidad Nacional de General Sarmiento, Los Polvorines, Buenos Aires, Argentina}
\affiliation{$^2$ National Scientific and Technical
Research Council, Argentina}
\affiliation{$^3$Max Planck Institute for Intelligent Systems,   Heisenbergstr.\ 3, 70569 Stuttgart,   Germany}
\affiliation{$^4$IV Institute for Theoretical Physics, University of Stuttgart, Pfaffenwaldring 57, 70569 Stuttgart, Germany}
\affiliation{$^5$  Helmholtz Institut Erlangen-N\"urnberg  for Renewable Energy (IEK-11), Forschungszentrum J\"ulich, F\"urther Str. 248, 90429, N\"urnberg, Germany}
\affiliation{$^6$ Departament de F\'isica de la Materia Condensada, Universitat de Barcelona, Barcelona, Spain}

\begin{abstract}
We report on the onset of anti-resonant behaviour of mass transport systems driven by time-dependent forces.  Anti-resonances arise from the coupling of a sufficiently high number of space-time modes of the force. The presence of forces having a wide space-time spectrum, a necessary condition for the formation of an anti-resonance, is typical of confined systems
with uneven and deformable walls that induce entropic forces dependent on space and time.  
We have analyzed, in particular, the case of polymer chains confined in a flexible channel and shown how they can be sorted and trapped.
The presence of resonance-antiresonance pairs found can be exploited to design protocols able to engineer optimal transport processes and to manipulate the dynamics of nano-objects.

\end{abstract}
\maketitle

\section{Introduction}
Many biological processes and micro- and nano-technological applications depend on the dynamics of soft systems, such as electrolytes, polymers and macromolecules~\cite{soft2013,Lin2016,RMP_Plants2016,Reisner_2012,Palyulin2014}, which are confined. This is, for example, the case of ionic and microfluidic channels~\cite{Albers_book,Lyderic_Charlaix}, blue energy rectifiers~\cite{Brogioli2009} and DNA-sequencing devices~\cite{Reisner_2012}.
Due to their small size, the dynamics of these systems are strongly affected by thermal noise and by noise generated by external sources.   

The interplay between noise and external forces has been so far analyzed for cases in which only one or some few spatial and/or temporal modes contribute to the force. Indeed, in Brownian ratchets, although the force may consist of different spatial modes, its dependence on time typically reduces to a single mode~\cite{Wu16,Graydon17,optical,Brox17,MagR19,Abbott19,Elecph1}.  
In contrast, in ghost stochastic resonance~\cite{GR2012} multiple temporal frequencies are involved in the force but just a single  spatial modes is excited. 
So far, the most general situation in which the force spectrum includes harmonics of a higher order both in time and space has not been analyzed. Such a scenario can be realized by means of optical, electric or magnetic fields ~\cite{Zemanek19,ElecR19, Magnet2} or by forces resulting from the geometrical confinement of the particles. This is 
the case of the ondulatory motion of worm-like organisms~\cite{wavy,RADTKE2017125}, peristaltic pumping~\cite{BKA,peris1969,peris1993,CR2017}, fluctuating ion channels and pores ~\cite{Cohen2011,Ikonen2012,Sarabadani2015,Mondal2016,Menais2016,Cecconi2017,Xu2018,Dubey2019,Du2020,Marbach2020,Marbach2019,ion} as well as synthetic soft micro- nanofluidic devices ~\cite{cells1,cells2,bact1}. Knowing what the impact of higher order harmonics is on the response of the system is a question of great current interest.

In this Article, we show that when a greater number of space-time modes contributes to the force spectrum, the particle current develops resonance-antiresonance pairs that increase or decrease the current and can therefore, by engineering the spectrum, set up optimal scenarios for the manipulation and transport of particles. The presence of antiresonances is detected from a perturbative analysis of the average particle current that progressively incorporates different modes into the force spectrum. 
The analytical predictions are supported by numerical solutions performed for the case of a single polymer chain confined in a channel whose walls undergo sustained oscillations. 
We show that if the amplitude of the oscillations of the effective force on the polymer is large enough, the current  exhibits a rich Fourier spectrum, even for simple single-mode oscillations of the shape of the walls. 
Our numerical data confirm that the onset of the resonance-antiresonance pairs is controlled by the "richness" of the Fourier  spectrum of the effective force acting on the polymer.

\section{Resonance-Antiresonance pair formation}

\begin{figure}
\centering
 \includegraphics[width=0.35\textwidth]{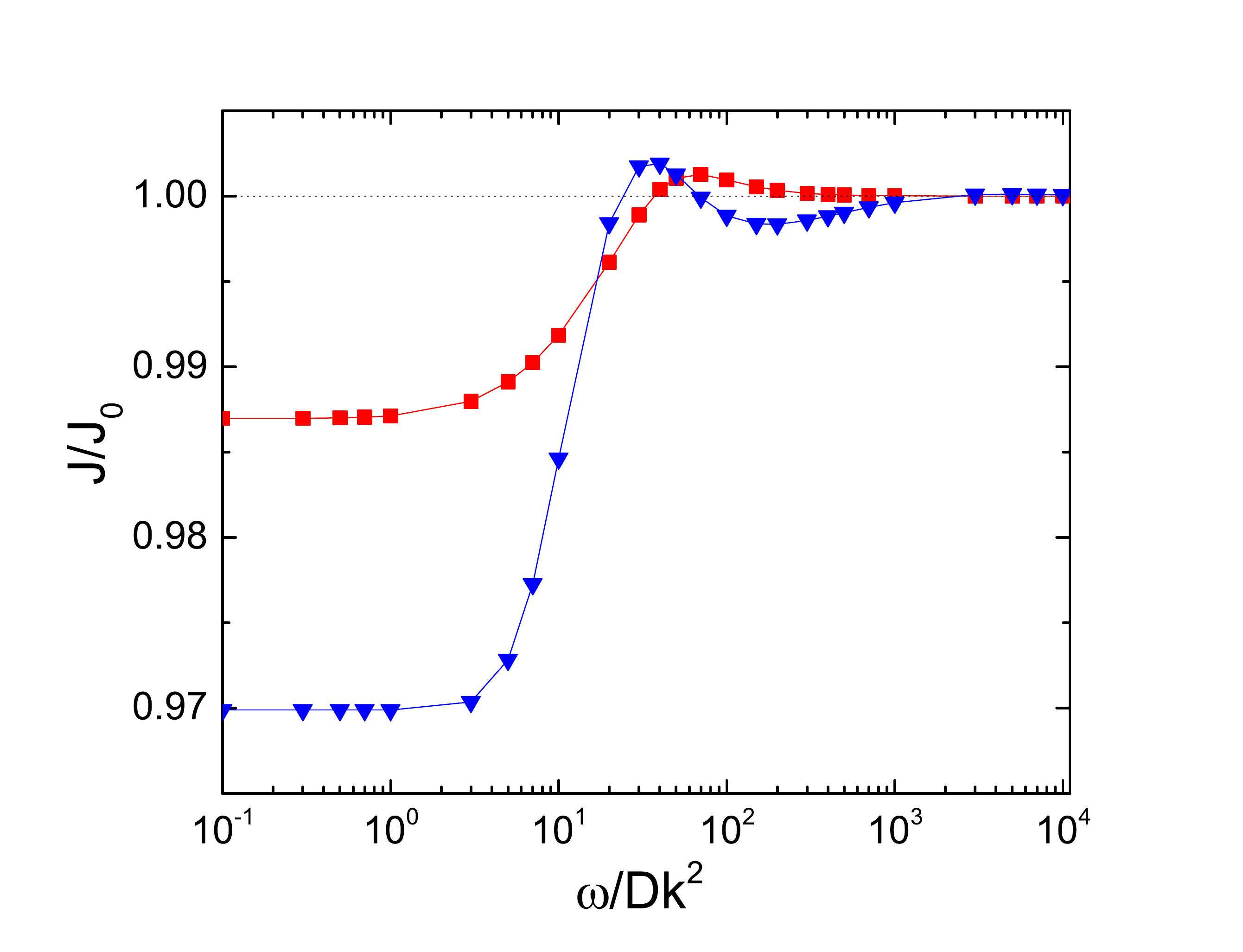}
\caption{Particle current J normalized to the current in a planar channel $J_{0}=\beta D \rho_{0} N f_0$ for $N_{\omega}=9$ temporal modes and  $N_k=1,12$ spatial modes corresponding  respectively to red squares and blue downward triangles.
$\omega$ and $k$ denote respectively, the frequency and wave number of the oscillatory channel.
Curves have been obtained for a force  with  $\text{Re}(f_{p,v})=\frac{8}{5}\frac{1}{p}$ and $ \text{Im}(f_{p,v})=\frac{12}{5}(-1)^{v}\frac{1}{p}$.}
\label{fig:J-theo}
\end{figure}

In the overdamped regime, the evolution of the probability distribution $\rho(x,t)$ is governed by the Smoluchowski equation 
\begin{align}
    \frac{\partial \rho(x,t)}{\partial t}=-\frac{\partial J(x,t)}{\partial x}=D\frac{\partial }{\partial x} \left[\frac{\partial }{\partial x} -\beta f(x,t)\right] \rho(x,t)
    \label{eq:smol}
\end{align}
where $D$ is the diffusion coefficient of the particle and $\beta^{-1}=k_BT$, with $k_B$ the Boltzmann constant and $T$ the absolute temperature.
After relaxation of the initial conditions, the density and current averaged over time (over the period of the force) are no longer dependent on time and reach a stable value. In order to obtain an analytical view of the process, we assume that the local force can be decomposed as 
\begin{align}
    f(x,t)=f_0+f_1(x,t)
\end{align}
with $f_{0}$ a constant force. We expand both the time averaged density and flux for small values of $f_1$ (for a detailed derivation see Appendix~\ref{app:analytics}). 
At first order, the time-averaged flux reads
\begin{equation}
    J^{(1)}= 2\sum_{p}\sum_v \bar{\rho}\beta ^2|f_{p,v}|^2\frac{\beta \bar{f}_0}{k_v^2}\frac{\Gamma_{p,v}-1}{\Gamma_{p,v}^2+4\left(\frac{\beta f_0}{k_v}\right)^2}
    \label{eq:J1}
\end{equation}
where $|f_{p,v}|$ is the amplitude of the Fourier mode of the force with frequency $\omega_p$ and wave number $k_{v}$ and
\begin{align}
    \Gamma_{p,v} =\left(\frac{\omega_p}{D k_v^2}\right)^2+1-\left(\beta\frac{ \bar{f}_0}{k_v}\right)^2
\end{align}
is a function of the ratios between the three relevant time scales: the diffusion time $\sim 1/(D k_v^2)$ for the mode with wavelength $2\pi/k_v$, the period of the force  $\sim 1/\omega_p$ and the drift time $\sim 2\pi/(\beta D f_0 k_v)$.
The second order contribution reads:
\begin{align}
 J^{(2)}= 2 \frac{\beta D}{L}\sum_p \sum_v \text{Re}(R^{(2)}_{p,v} f^*_{p,v})   
 \label{eq:J_2}
\end{align}
where $R^{(2)}_{p,v}$ are the amplitudes of the modes of the probability density at second order in the expansion (see Appendix~\ref{app:analytics}).
\begin{eqnarray}
R^{(2)}_{p,v}=-\frac{1}{\Gamma_{p,v}+2\iota \dfrac{\beta \bar{f}_0}{k_v}} \mathlarger{\sum_{n\neq p}}\hspace{0.1cm} \mathlarger{\sum_{i}} R^{(1)}_{n,i}\hspace{0.1cm} \dfrac{\beta f_{p-n,v-i}}{k_v} \nonumber \\
 \left[2\iota -\iota \dfrac{k_{v-i}}{k_v}\dfrac{k_{v-i}-2k_i}{k_v}-3\left(\dfrac{\beta \bar{f}_0}{k_v}\right)^2\right]
\label{eq:R2}
\end{eqnarray}
with 
\begin{align}
    R^{(1)}_{n,i}=- \dfrac{\beta f_{n,i}}{k_i} \bar{\rho}\frac{\iota-\frac{\beta \bar{f}_0}{k_i}}{\Gamma_{n,i}+2\iota\frac{\beta \bar{f}_0}{k_i}}\,.
\end{align}
For just one spatial or temporal mode, $R^{(2)}_{p,v}$ is zero since either $R^{(1)}_{p,v}$ or $f_{p,v}$ in Eq.~\eqref{eq:R2} are zero and hence the contribution of $J^{(2)}$ to $J$ becomes negligible. Accordingly, the total current is determined solely by its first order contribution $J^{(1)}$.
Fig.\ref{fig:J-theo}  shows that in this case $J$ exhibits a resonant peak in agreement with previous studies on polymer translocation across fluctuating pores~\cite{Cohen2011,Ikonen2012,Sarabadani2015,Mondal2016,Cecconi2017}. 
When multiple spatial and temporal modes are present, 
 $R^{(2)}_{p,v}\neq 0$ and hence $J^{(2)}\neq 0$.
The coupling of these modes, not present in $J^{(1)}$, causes the formation of the resonant-antiresonant doublet of the current shown in Fig.\ref{fig:J-theo}. \\

\section{
Resonance-antiresonance doublets for particle sorting and trapping}
\begin{figure}
\includegraphics[width=0.35\textwidth]{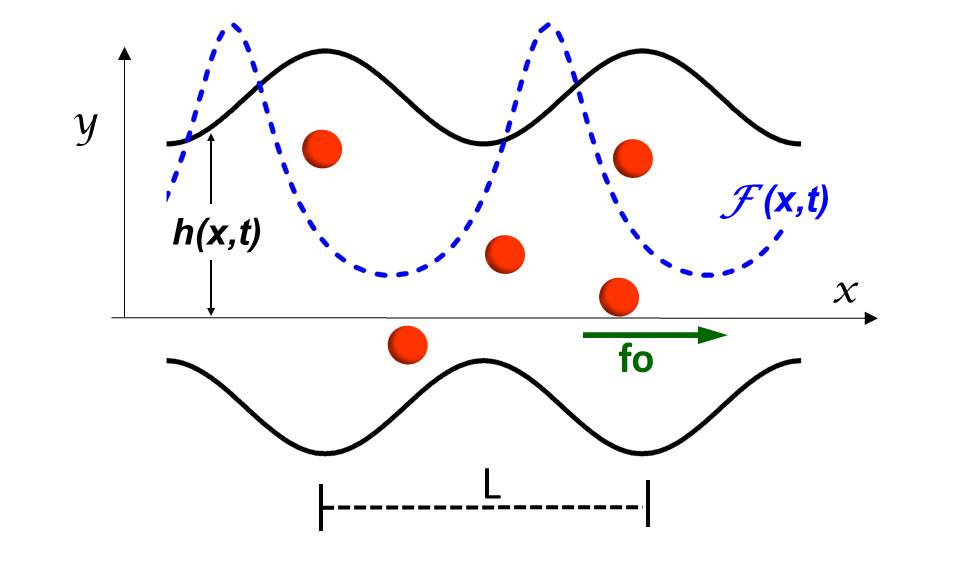}
\caption{Schematic representation of the Brownian motion of  microscopic objects in an oscillatory channel of period $L$ with a time-dependent radius $h(x,t)$ (in black). The free energy $\mathcal{F}(x,t)$ (in blue) consists of energetic and entropic contributions related to the applied force $f_{0}$ and to the entropic forces resulting from constrictions of the channel that give rise to a time- and space-dependent cross-sectional area.}
\label{fig:Figura2}
\end{figure}

A typical case in which the spatio-temporal spectrum of the external force induces an antiresonance is that of a nano-object moving in an environment bounded by irregular walls whose shape changes over time (Fig.\ref{fig:Figura2}).
In the case of a flexible pore, even though the deformation, characterized by its half-height, is given in terms of a single spatial and temporal mode
\begin{align}
    h(x,t)=h_0+h_1\cos(kx)\sin(\omega t)
    \label{eq:h_simple}
\end{align}
the effective force induced on the confined particle is characterized by a rich Fourier spectrum. In order to show this, in  the following we specialize the case of a Gaussian polymer chain confined in a 3D flexible channel whose half-height undergoes periodic oscillations according to Eq.~\eqref{eq:h_simple}.

When $h(x,t)$ varies smoothly (|$\partial_x h(x,t)| \ll 1$), the dynamics of the polymer can be mapped onto that of a point-like particle in the presence of the effective local free energy~\cite{bianco}
\begin{align}
\beta\mathcal{F}(x,t) = -\beta N f_0 x + \beta A(x,t)
\label{eq:def-F}
\end{align}
with
\begin{align}
&\beta A(x,t) = \beta A_{cm}(x,t)+\beta A_{mon}(x,t)\\
&\equiv -2 \ln \left[\frac{16 h(x,t)}{h_0 \pi^2}\right]-2\ln \left[\sum^{\infty}_{p=odd}\!\!\!\! \frac{e^{-\pi^2 p^2 \left(\frac{R_G}{2 h(x,t)} \right)^2}}{p^2}  \right]\nonumber
\label{eq:def-A}
\end{align}
where $R_{G}=\sqrt{\frac{N r^{2}}{6}}$ is the radius of gyration of the polymer in an unbounded medium, $N$ the number of monomers and $r$ the Kuhn length. The first term on the rhs of Eq.~\eqref{eq:def-F} is an energetic contribution due to the action of external force $f_0$ on each monomer. The second term, denoted as $\beta A_{cm}(x,t)$, is an entropic contribution resulting from the number of allowed positions (microstates) of the centre of mass of the polymer within the channel~\cite{Zwanzig,rubi2001b,Burada2007,Malgaretti2013,Malgaretti2021} whereas the third one $\beta A_{mon}(x,t)$ accounts for the correction to the second term due to the presence of the other monomers~\cite{bianco}. 
\begin{figure}
\centering
\includegraphics[width=0.45\textwidth]{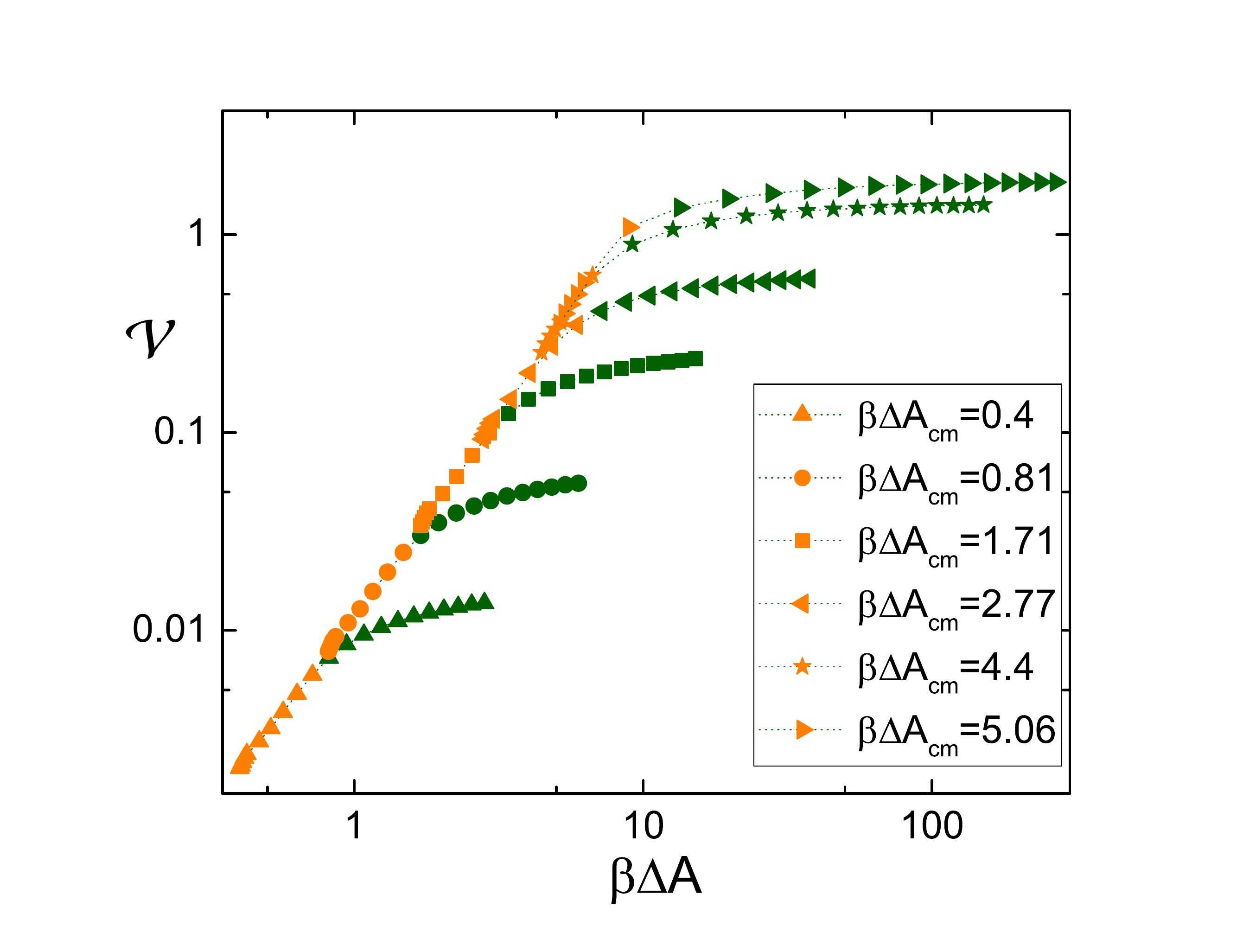}
\caption{Variance $\mathcal{V}$  versus $\beta \Delta A$. Orange symbols correspond to the case $\Delta A_{mon}/\Delta A_{cm}<1$ whereas green symbols to $\Delta A_{mon}/\Delta A_{cm}>1$. Averages are taken during a time $T=2\pi/\omega$.}
\label{fig:variance}
\end{figure}
The diffusion current $J$ in Eq.~\eqref{eq:smol} is obtained by considering that the force is $f(x,t)= -\frac{\partial \mathcal{F}(x,t)}{\partial x}$ and  that,  according to the Rouse model,
$D=D_0/N$ is the diffusion coefficient of the polymer with $D_0$ that of a single monomer. 

We remark the fact that even in the present case in which channel oscillations consist of a single $k$-$\omega$ mode, the spectrum of the force $f(x,t)$ contains contributions from higher order spatial and temporal modes whose amplitudes depend on the maximum free energy difference
\begin{align}
    \beta \Delta A =\max_{x,t} \beta A(x,t)-\min_{x,t}\beta A(x,t)\,.
\end{align}
For $\beta \Delta A\neq 0$, the amplitude of the modes is sensitive to the nature of the system.
For confined hard spheres, $A_{mon}$ drops out and the value of $\Delta A$ is controlled only by the geometry of the channel. Typical experimental realizations limits this value to $\beta \Delta A \lesssim 3$~\cite{dinsmore1999}.
For confined uncharged polymers, $A_{mon}$ can give rise to arbitrary large values of $\Delta A$ upon increasing the radius of gyration $R_G$. Finally, for charged colloids, polymers or even more complex particles the magnitude of $\Delta A$ can be quite large ~\cite{Malgaretti2013,Leith2016}
and is controlled by both geometry of the channels and (electrostatic) interactions with the channel walls. 

Our analysis reveals that the key factor for the onset of an antiresonance is the existence of a "rich" spectrum of both spatial and temporal modes of the external force. The "spectral richness" of the power spectrum $\mathcal{P}_{p,v}=f^2_{p,v}$ of a given external force $f(x,t)$ is captured by means of the variance
\begin{align}
 \mathcal{V}\equiv\left\langle \mathcal{P}_{p,v}^2\right\rangle-\left\langle\mathcal{P}_{p,v}\right\rangle^2   
\end{align}
where $\left\langle...\right\rangle$ denotes average over $p$ and $v$.
Fig.\ref{fig:variance} shows $\mathcal{V}$ versus $\beta \Delta A$, for different values of $\beta \Delta A_{cm}$ and $\beta \Delta A_{mon}$. For a given $\Delta A$ and $ \Delta A_{mon}/\Delta A_{cm} < 1$, the variance displays a power law behavior represented by the orange master branch: $\mathcal{V} \approx (\beta \Delta A)^{2}$. For $\Delta A_{mon}/\Delta A_{cm} > 1$, it shows a significant sensitivity to changes of the parameters that characterize the channel geometry and the polymer, as evidenced by the appearance of multiple secondary branches (green symbols).
On the other hand, given a value of $\Delta A_{cm}$ and varying $\Delta A$ along the master branch towards the corresponding secondary branch, the variance increases. 
\begin{figure}
\centering
\subfloat[][]{\includegraphics[width=0.45\textwidth]{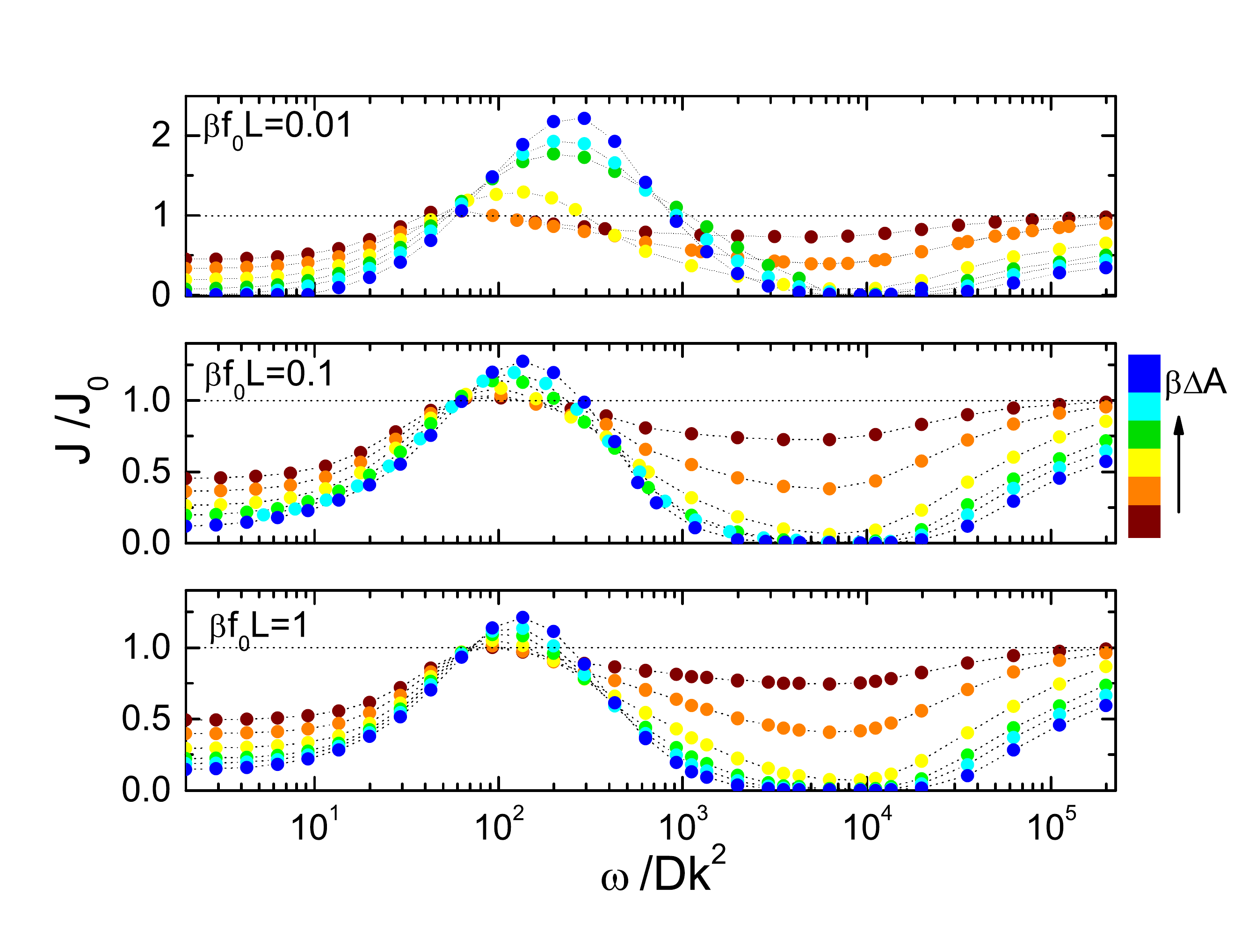}}\vspace{0.01cm}
\subfloat[][]{\includegraphics[width=0.4\textwidth]{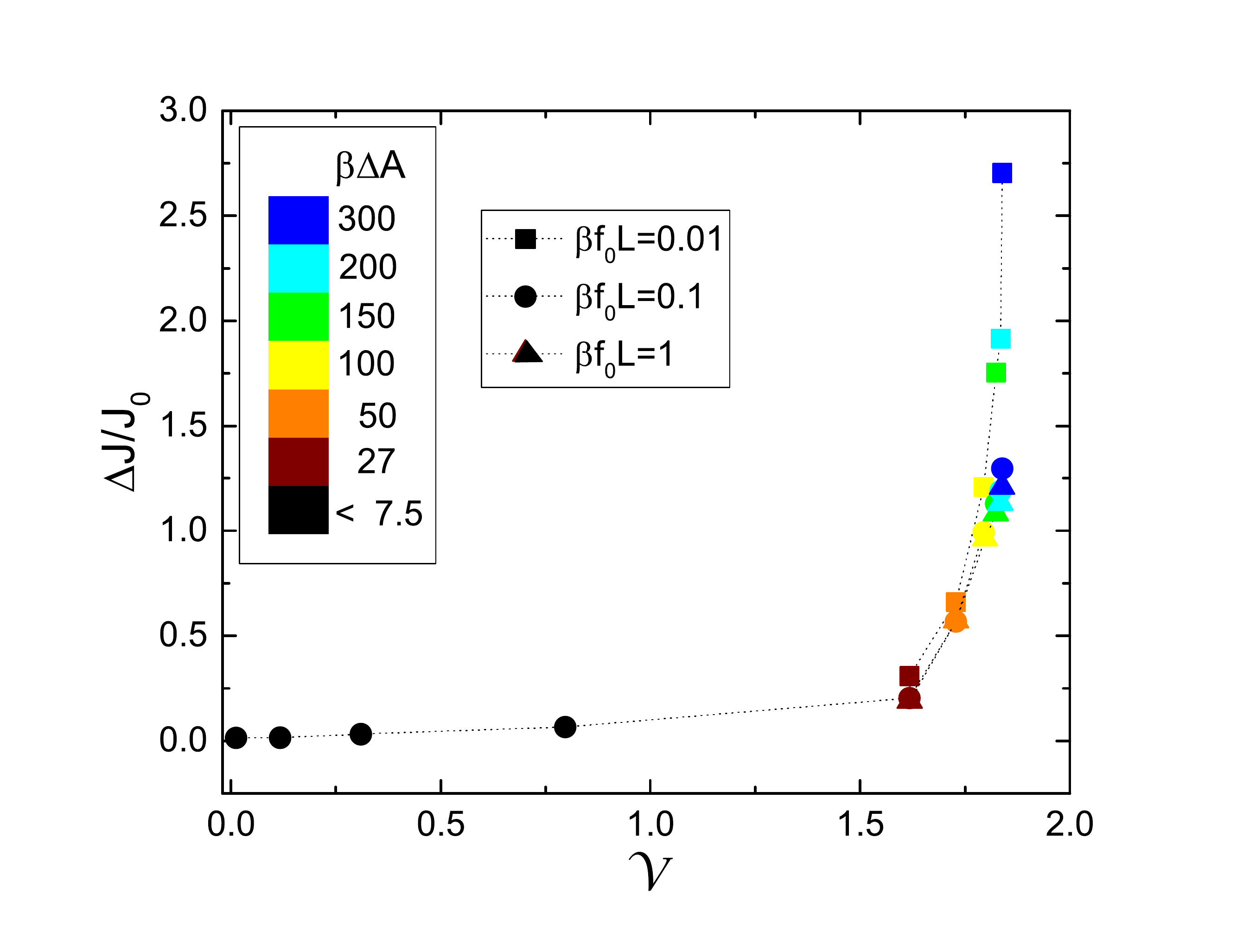}}
\caption{(a): $J/J_{0}$ as a function of $\omega/(D k^2)$ for $\beta f_0 L=0.01, 0.1, 1$, $R_G=0.03, N=10$ and $\beta\Delta A_{cm} \approx 5$. Different colors correspond to increasing values (from brown to blue) of $\beta\Delta A = 27, 50, 100, 150, 200, 300$. The reference current is $J_{0}= \beta D \rho_0 N f_0$, the current in the case of point particles moving through a flat channel. (b): $\Delta J/J_0$ (See Eq.~\eqref{eq:DJ}) versus $\mathcal{V}$ for different values of the force and $\beta \Delta A$ (see legend).}
\label{comparo}
\end{figure}

The  presence of higher order harmonics of the force originates the antiresonance in the particle current $J$, normalized to the current for a flat channel $J_{0}$, plotted in Fig.\ref{comparo}-a. The different colors stand for the upward series of values $\beta \Delta A$
at which $\mathcal{V}$ grows. In all cases, resonant/antiresonant maximum/minimum emerge with more pronounced minima at  higher values of $\mathcal{V}$.
Fig.\ref{comparo}-a also shows the presence of different transport regimes.
The limit $\omega/D k^2 \rightarrow 0$ corresponds to the dynamics of a point-like particle moving in an almost static effective potential characterized by a free energy barrier $\beta\Delta A$ and under the action of an external force $f_0$.
In contrast, for $\omega/D k^2 \rightarrow \infty$ the walls of the channel oscillate so fast, as compared to the typical time scale of the polymer, that it experiences an average potential equivalent to that of a polymer moving through a flat channel under the action of the external force $f_0$~\footnote{
The dynamics of the polymer can be mapped onto that of a point-like particle with the effective free energy $\beta A(x,t)$, provided that the relaxation time scale of the plolymer, $\tau=2R^2_G/\pi^2 D$, is much faster than any other time scale}. 
In between these two limiting regimes, the current shows a non-monotonic behavior with a resonance and an antiresonance. In particular, for all values of $\beta \Delta A$ , $J$ exhibits a resonant peak at $\frac{\omega}{D k^2}\simeq 100$, as predicted by our model (see Eq.~\eqref{eq:J1}).
Interestingly, although the analytical results are obtained via a perturbative approach, the prediction about the appearance of a resonant peak is robust for high barriers and its location depends only weakly on the magnitude of $\beta \Delta A$.  However, its amplification for increasing values of $\beta \Delta A$ is not found under this approach. 
The analytical results also predict the appearance of the antiresonance observed in Fig.\ref{comparo}a). 
The onset of the antiresonance minimum is captured by the quantity
\begin{equation}
    \Delta J\equiv J\left(\frac{\omega}{Dk^2}=10^2\right)-J\left(\frac{\omega}{Dk^2}=10^4\right)
    \label{eq:DJ}
\end{equation}
Indeed, in the absence of an antiresonant minimum $\Delta J/J_0\ll 1$, whereas once the antiresonance sets $\Delta J/J_0\gtrsim 1$.
Fig.\ref{comparo}b shows that for a "simple" Fourier spectra ($\mathcal{V}<1.5$) $\Delta J/J_0\ll 1$, whereas for larger values of $\mathcal{V}$ i.e. for a "rich" Fourier spectra  $\Delta J/J_0\gtrsim 1$ which triggers the antiresonance. 
\begin{figure}[t]
\centering
\subfloat[][]{\includegraphics[scale=0.15]{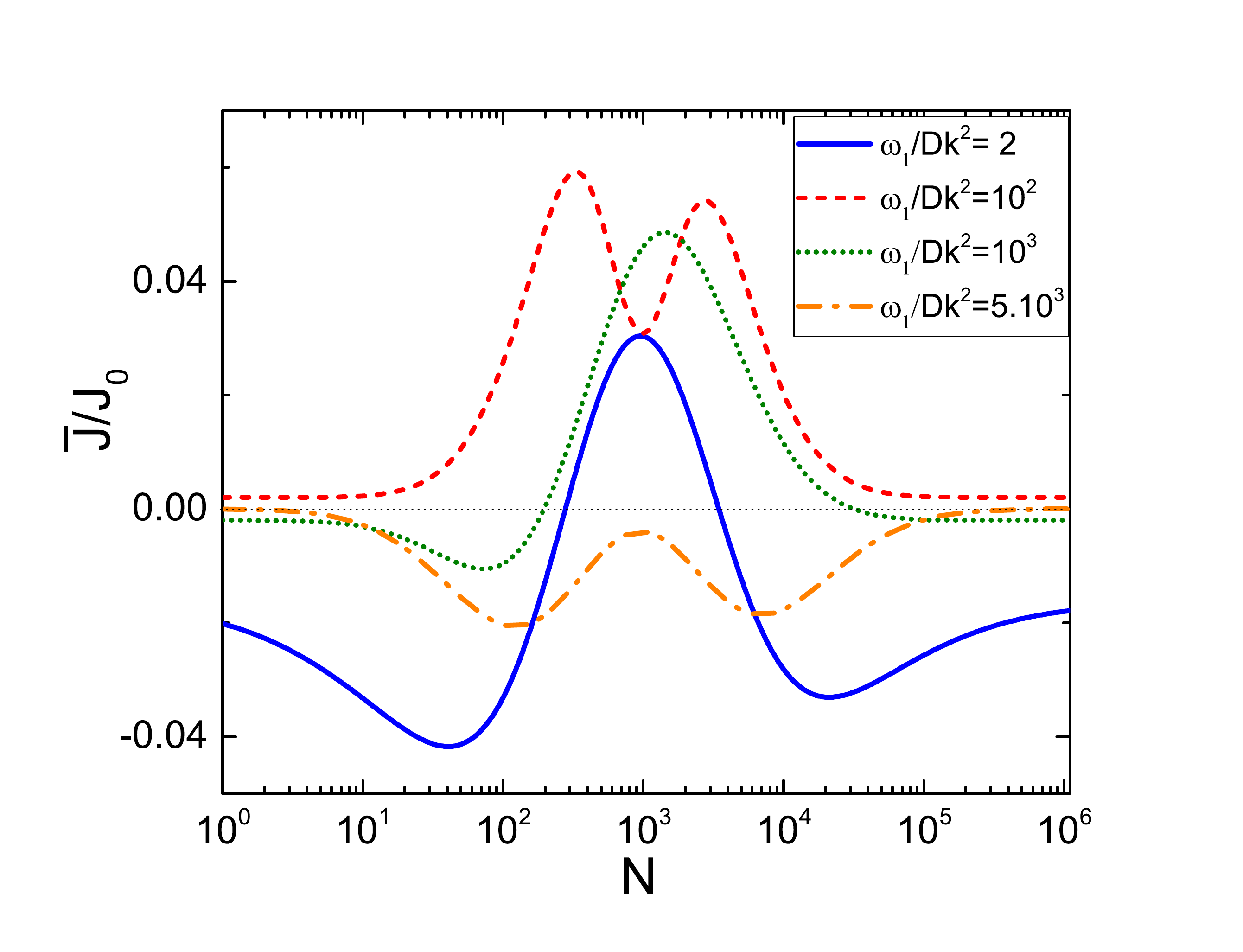}}
\subfloat[][]{\includegraphics[scale=0.15]{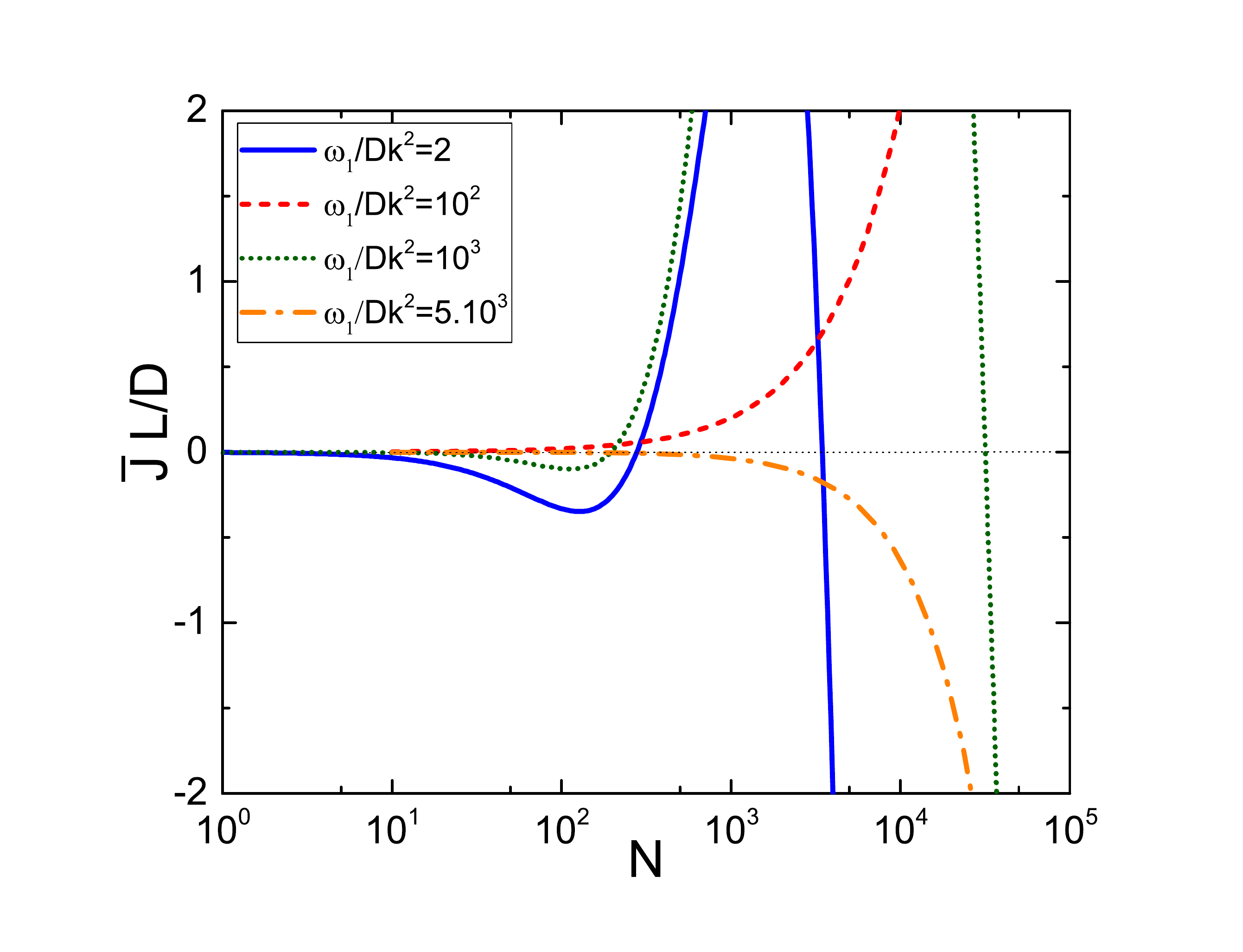}}
\caption{(a) $\bar{J}/J_{0}$ and (b) $\bar{J}L/D$ versus $N$ for the proposed protocol. $\omega_1=2$ (\textcolor{blue}{full}), $\omega_1=10^2$ (\textcolor{red}{dashed}), $\omega_1=10^3$ (\textcolor{OliveGreen}{dotted}), $\omega_1=5\cdot10^3$ (\textcolor{Orange}{dotted-dashed}) and $\omega_2=3\cdot10^4$ in all cases. Other parameters: $\beta \Delta A_{cm}=5$, $\beta f_{0}L=0.1$, $b/h_0=0.01$.
}
\label{fig:JRG}
\end{figure}
The presence of a resonance-antiresonance doublet can be exploited to design a dynamical protocol able to sort and trap polymers. The protocol runs as follows: an external force $f_0$ is applied along with a time modulation of the channel half-height $h(x,t)$ characterized by a frequency $\omega_1$ during a time $T/2$.  Then the force is switched to $-f_0$ and the frequency to $\omega_2$. It is also applied during a time $T/2$. The time $T$ is chosen such that $T\gg 2 \pi /\omega_{1,2}$. Finally, we calculate the mean current in T 
\begin{align}
    \bar{J}=\frac{J_1 + J_2}{2}\,,
\end{align}
where $J_i$ are the average currents in each half-period computed as indicated previously.

Fig.\ref{fig:JRG}a, shows $\bar{J}$ as function of $N$.
We observe that chains with lengths in the range $10^2$ to $10^4$ times a monomer size can be immobilized and separated with a good resolution.
The choice of $\omega_{1,2}$ is crucial to improve the performance of the protocol that requires the presence of a deep antiresonant minima that 
occur for large values of $\mathcal{V}$. Optimal conditions are achieved when one of the frequencies, say $\omega_2$, is the antiresonant one (Fig.\ref{fig:JRG}a).
\noindent 
The protocol is particularly effective for long polymers, for which $\bar{J}$ crosses the zero value with a very steep profile (see Fig.\ref{fig:JRG}b), with a value much larger than the characteristic diffusion velocity $D/L$. These conditions essentially take place when $\omega_1$ is out of the amplification regime (resonance). For polymers whose lengths are very similar to that of the target (polymer with $\bar{J}=0$), the signal-to-noise ratio is small which gives rise to a poor trapping resolution.

\section{Conclusions}
In this article, we have shown that cooperation between the noise and an external forces whose spectrum contains a high enough number of harmonics leads to the appearance of new resonant states of the system. 
The occurrence of the resonant peak results from the well-known synchronisation of the hopping events induced by noise with the oscillations of the force (in our case of entropic origin).
The increase in the number of modes of the 
force causes the system to move from a regime of
a single resonant peak, typical of the case of a small
number of modes, to a regime in which a pair of resonant-antiresonant
peaks arises. The presence of a large number of modes causes the noise to synchronise with the different modes, which can lead to destructive interferences that make the particle current to decrease. Forces whose spectrum consists
of a high number of coupled modes required for the
appearance of antiresonances arise naturally in the case
of systems confined by uneven and flexible walls.

We have shown this feature by analyzing the dynamics of a polymer chain confined within a channel whose walls undergo sustained oscillations. Our numerical data confirm the predictions of the analytical model and show a clear correlation between the number of modes (encoded in the variance of the Fourier power spectrum $\mathcal{V}$) and the onset of the antiresonant minimum. 

The results obtained broaden the understanding of the resonant phenomena in drift-diffusion systems and may provide the rational for the interpretation of dataset existing  in the literature~\cite{ion,Kadem17,ElecR19,Bauerle20}.
The existence of the new regimes we have found, in which resonances and antiresonances coexist, opens up the possibility of designing the driving force to be able to control the flow of particles and to manipulate nano-objects in order to study their properties. 

M.F.C. thanks PIO-CONICET and PICT 18/2036, J.M.R. wants to thank financial support of MICIU of  the Spanish Government, under Grant No. PGC2018-098373-B-I00.
\appendix
\section*{\label{app:analytics} Appendix}
Our starting point is the $1D$ advection-diffusion equation:
\begin{equation}
\dot{\rho}(x,t)=\partial_{x}\left[D\partial_{x}\rho(x,t)-D\beta\rho(x,t)f(x,t)\right]
\label{eq:smol-1}
\end{equation}
where $f(x,t)$ is the effective force acting
on the point-like particle (for conservative forces we have $f(x,t)=-\partial_{x}W(x,t)$).
We are interested in the long-time regime in which $\rho(x,t)$ can be
assumed to be periodic in time. Accordingly, we represent both $\rho$ and $f$ as:
\begin{subequations}\label{eq:four-T}
\begin{eqnarray}
\rho(x,t) & = & \sum_{n=-\infty}^{\infty}\rho_{n}(x)e^{\iota\omega_{n}t}\\
&\Rightarrow & \rho_{n}(x)=\frac{1}{2\pi}\int\rho(x,t)e^{-\iota\omega_{n}t}dt\\
f(x,t) & = & \sum_{n=-\infty}^{\infty}f_{n}(x)e^{\iota\omega_{n}t}\\
&\Rightarrow & f_{n}(x)=\frac{1}{2\pi}\int f(x,t)e^{-\iota\omega_{n}t}dt
\end{eqnarray}
\end{subequations}
The real nature of both $\rho$ and $f$ imposes the  constraints on the amplitudes
of the modes: 
\begin{subequations}\label{eq:four-real}
\begin{eqnarray}
\text{{Re}\ensuremath{\left(\rho_{n}\right)}} & = & \text{{Re}\ensuremath{\left(\rho_{-n}\right)}}\\
\text{{Im}\ensuremath{\left(\rho_{n}\right)}} & = & \text{-{Im}\ensuremath{\left(\rho_{-n}\right)}}\\
\text{{Re}\ensuremath{\left(f_{n}\right)}} & = & \text{{Re}\ensuremath{\left(f_{-n}\right)}}\\
\text{{Im}\ensuremath{\left(f_{n}\right)}} & = & \text{-{Im}\ensuremath{\left(f_{-n}\right)}}
\end{eqnarray}
\end{subequations}
Using Eqs.~\eqref{eq:four-T}, Eq.~\eqref{eq:smol-1} one obtains
\begin{equation}
    \begin{split}
&\sum_{n=-\infty}^{\infty}\iota\omega_{n}\rho_{n}(x)e^{\iota\omega_{n}t}=\\
&\sum_{n,n=-\infty}^{\infty}\partial_{x}\left[D\partial_{x}\rho_{n}(x)-D\beta\rho_{n}(x)f_{m}(x)e^{\iota\omega_{m}t}\right]e^{\iota\omega_{n}t}
\end{split}
\label{eq:fuorier-tot}
\end{equation}
Multiplying both sides of Eq.\eqref{eq:fuorier-tot} by $e^{-\iota\omega_{p}t}$ and
integrating in $t$ we obtain:
\begin{equation}
\iota\omega_{p}\rho_{p}(x)=D\partial_{x}\left[\partial_{x}\rho_{p}(x)-\sum_{n=-\infty}^{\infty}\beta\rho_{n}(x)f_{p-n}(x)\right]
\label{eq:fuorier-tot-1}
\end{equation}
We assume periodic boundary conditions: 
\begin{eqnarray}
\rho(0,t) = \rho(L,t)
\end{eqnarray}
and use mass conservation 
\begin{eqnarray}
\int_{0}^{L}\rho(x,t)dx  = L\bar{\rho}\,,\quad \forall t
\end{eqnarray}
In Fourier representation these equations leads to 
\begin{eqnarray}
\rho_{0}(0) & = & \rho_{0}(L)\label{eq:BC-1}
\end{eqnarray}
and 
\begin{eqnarray}
\rho_{n}(0) & = & \rho_{n}(L)\label{eq:BC-2}
\end{eqnarray}
Since mass conservation holds at all times, we have
\begin{eqnarray}
\int_{0}^{L}\rho_{0}(x)dx & = & L\bar{\rho}\\
\int_{0}^{L}\rho_{n}(x)+\rho_{-n}(x)dx & = & 0,\,\forall n\neq0
\end{eqnarray}
In the long-time limit, in which the initial
conditions have relaxed, integrating over a time period removes any time dependence and leads to the steady time-averaged
density probability and flux. Using the expression of the current
$j(x,t)=-\left[D\partial_{x}\rho(x,t)-D\beta\rho(x,t)f(x,t)\right]$,
we obtain : 
\begin{eqnarray}
0 & = & \frac{1}{T}\int\limits _{t_{0}}^{t_{0}+T}\partial_{x}j(x,t)dt
\end{eqnarray}
From this expression, one obtains
\begin{align}
J  &=\frac{1}{T}\int\limits _{t_{0}}^{t_{0}+T}j(x,t)dt \\ &=-D\left[\partial_{x}\rho_{0}(x)-\sum_{n=-\infty}^{\infty}\beta\rho_{n}(x)f_{-n}(x)\right]
\end{align}
Notice that for $1D$ systems at steady state, $J$ does not depend on position.
Using the real nature of both $\rho$ and $f$, we obtain:
\begin{equation}
\begin{split}
J  &=  -D[\partial_{x}\rho_{0}(x)-\beta\rho_{0}(x)f_{0}(x)-\\
& 2\sum_{n=1}^{\infty}\beta\left[\text{Re}\left(\rho_{n}(x)\right)\text{Re}\left(f_{n}(x)\right)+\text{Im}\left(\rho_{n}(x)\right)\text{Im}\left(f_{n}(x)\right)\right]]
\end{split}
\end{equation}
Since $J$ does not depend on space, we can integrate it in $x$ to obtain
\begin{equation}
\begin{split}
J  &=  -\frac{D}{L}\int_0^L[\partial_{x}\rho_{0}(x)-\beta\rho_{0}(x)f_{0}(x)-\\
 & 2 D\sum_{n=1}^{\infty}\beta\left[\text{Re}\left(\rho_{n}(x)\right)\text{Re}\left(f_{n}(x)\right)+
\text{Im}\left(\rho_{n}(x)\right)\text{Im}\left(f_{n}(x)\right)\right]]dx
\end{split}
\end{equation}
which due to the periodic boundary conditions reads:
\begin{equation}
\begin{split}
J  &=  \frac{D}{L}\int_0^L\beta\rho_{0}(x)f_{0}(x)+\\
& 2\sum_{n=1}^{\infty}\beta\left[\text{Re}\left(\rho_{n}(x)\right)\text{Re}\left(f_{n}(x)\right)+\text{Im}\left(\rho_{n}(x)\right)\text{Im}\left(f_{n}(x)\right)\right]dx
\end{split}
\end{equation}
Using now Eqs.~\eqref{eq:four-real}, the last expression can be rewritten as
\begin{equation}
\begin{split}
J  &=  \frac{D}{L}\int_0^L\beta\rho_{0}(x)f_{0}(x)+2\sum_{n=1}^{\infty}\beta \text{Re}\left(\rho_{n}(x)f^*_{n}(x)\right)dx
\end{split}
\label{eq:J-app}
\end{equation}
To simplify the calculation, and without loss of generality, we choose
\begin{align}
     \text{Im}(f_i(x))=0
     \label{eq:alt-1}
\end{align}
Accordingly, using
\begin{align}
    \rho_n(x)=\text{Re}(\rho_n(x))+\iota\text{Im}(\rho_n(x))\equiv R_n(x)+\iota I_n(x)
\end{align}
Eq.\eqref{eq:fuorier-tot-1} can be rewritten as:
\begin{equation}
\begin{split}    \omega_p R_p(x)&=D\partial_x\left[\partial_x I_p(x)-\sum_n \beta I_n f_{p-n}(x)\right]\\ 
    \omega_p I_p(x)&=-D\partial_x\left[\partial_x R_p(x)-\sum_n \beta R_n f_{p-n}(x)\right]
\end{split}
\end{equation}from which we get:
\begin{equation}
\begin{split}
I_p(x)& =-\frac{D}{\omega_p}\partial_x\left[\partial_x R_p(x)-\sum_n \beta R_n(x) f_{p-n}(x)\right]
\end{split}
\end{equation}
and
\begin{equation}
\begin{split}
\omega^2_p R_p(x)=&-D^2[\partial^4_x R_p(x)
    -2\partial_x^3\sum_n\beta R_n(x)f_{p-n}(x) \\
  &  +\beta\partial_x\sum_n [R_n(x)\partial_x^2 f_{p-n}(x)\\
  & +2(\partial_xR_n(x))(\partial_x f_{p-n}(x))]\\
   & +\partial_x\sum_n\beta^2\sum_m f_{p-n}(x)\partial_x (R_m(x)f_{n-m}(x))]
\end{split}
\label{eq:4-order}
\end{equation}
Eq.~\eqref{eq:4-order} is a fourth order ODE with non-constant coefficients. In order to get an analytical solution, we assume that the force is homogeneous:
\begin{align}
    f_0(x)=\bar{f}_0
    \label{eq:frc-simpl}
\end{align}
and that the amplitude of the higher modes, $f_q$ is small as compared to $f_0$ i.e.,  $f_q\ll f_0\,\forall q$. 
Under these assumptions, we can expand the equation for small values of $f_q$.
The zeroth order of Eq.~\eqref{eq:4-order} then reads:
\begin{equation}
\begin{split}
\omega^2_p R^{(0)}_p(x)= & -D^2 [\partial^4_x R^{(0)}_p(x)-2\beta f_0\partial_x^3  R^{(0)}_p(x)\\
& + \beta^2f_0^2\partial^2_x R^{(0)}_{p}(x)]
\end{split}
\label{eq:4-order-0}
\end{equation}
whose solution is:
\begin{align}
    R^{(0)}_0&=\bar{\rho}\\
    R^{(0)}_p&=0\quad \forall p\neq 0
\end{align}
At first order, we have 
\begin{equation}
\begin{split}
\omega^2_p R^{(1)}_p(x)&= -D^2 [\partial^4_x R^{(1)}_p(x)\\
& -2\beta \partial_x^3 [f_0  R^{(1)}_p(x)+\bar{\rho}f_p(x)]\\
& + \beta\bar{\rho}\partial^3_x f_p(x)+
 \beta^2f_0\bar{\rho}\partial^2_x f_{p}(x)]
\end{split}
\label{eq:4-order-1}
\end{equation}
We then use the space-Fourier representation
\begin{align}
    R^{(1)}_p(x)=\sum_n R^{(1)}_{p,n} e^{\iota k_n x}\\
    f_p(x)=\sum_n f_{p,n} e^{\iota k_n x}
\end{align}
After projecting onto the mode $n=v$, Eq.~\eqref{eq:4-order} reads:
\begin{equation}
\begin{split}
\left(\frac{\omega_p}{D k_v^2}\right)^2 R^{(1)}_{p,v} =& -[R^{(1)}_{p,v}+\iota \left(\bar{\rho}\frac{\beta f_{p,v}}{k_v}+2\frac{\beta f_0}{k_v} R^{(1)}_{p,v}\right)\\
 & -\frac{\beta f_0}{k_v}\left(\bar{\rho} \frac{\beta f_{p,v}}{k_v}+\frac{\beta f_0}{k_v} R^{(1)}_{p,v}\right)]
\end{split}
\end{equation}
whose solution is:
\begin{align}
   R^{(1)}_{p,v}&=- \dfrac{\beta f_{p,v}}{k_v} \bar{\rho}\frac{\iota-\frac{\beta f_0}{k_v}}{\Gamma_{p,v}+2\iota\frac{\beta f_0}{k_v}}\\
   \Gamma_{p,v}&=\left(\frac{\omega_p}{D k_v^2}\right)^2+1-\left(\beta\frac{ f_0}{k_v}\right)^2\label{eq:app-Gamma}
\end{align}
The second order gives:
\begin{equation}
\begin{split}
 -\frac{\omega^2_p}{D^2} R^{(2)}_p(x)&=\partial^4_x R^{(2)}_p(x)\\
& - \beta \partial_x^3 [R^{(2)}_{p}(x) \bar{f}_{0}+\sum_{n\neq p}R^{(1)}_{n}(x) f_{p-n}(x)]\\
    &+\beta\partial_x\sum_{n\neq p}[R^{(1)}_{n}(x)\partial_x^2 f_{p-n}(x)\\
    & +2\left(\partial_x R^{(1)}_{n}(x)\right)\left(\partial_x f_{p-n}(x)\right)]\\
    &+\beta^2 \bar{\rho}\sum_{n\neq p}\sum_{m\neq p} \partial_x\left[ f_{p-n}(x)\partial_x f_{n-m}(x)\right]\\
    &+ \beta^2 \bar{f}_0 \partial_x [\sum_{n\neq p} f_{p-n}(x)\partial_x R^{(1)}_{n}(x)\\
    & +\sum_{m\neq p} \partial_x\left(f_{p-m}(x) R^{(1)}_{m}(x)\right) ]\\
    &+\beta^2 \bar{f}_0^2\partial_x^2 R^{(2)}_p(x)
\end{split}
\label{eq:4-ord-2}
\end{equation}
Using 
\begin{equation}
    R^{(2)}_p(x)=\sum_i R^{(2)}_{p,i} e^{\iota k_i x}
\end{equation}
once projected, Eq.~\ref{eq:4-ord-2} becomes
\begin{equation}
\begin{split}
 -\frac{\omega^2_q}{D^2} R^{(2)}_{p,v}&= k^4_v R^{(2)}_{p,v}\\
& +2\iota\beta \sum_{n \neq p} \sum_i \left(k_i+k_{v-i}\right)^3 R^{(1)}_{n,i} f_{p-n,v-i}\\
& +2\iota \beta \bar{f}_0 k_v^3 R^{(2)}_{p,v}\nonumber\\
     &-\iota \beta \sum_{n\neq p}\sum_i [ k^2_{v-i}(k_i+k_{v-i})R^{(1)}_{n,i} f_{p-n,v-i}\\
     & -2 k_i k_{v-i}(k_i+k_{v-i}) R^{(1)}_{n,i} f_{p-n,v-i}]\nonumber\\
     &-\beta^2 \bar{\rho}\sum_{n\neq p}\sum_{m\neq p} [\sum_i   k_{v-i}(k_i+k_{v-i})f_{p-n,i} f_{n-m,v-i}]\nonumber\\
     &-\beta^2 \bar{f}_0 [2\sum_{n\neq p} \sum_i k_i(k_i+k_{v-i})f_{p-n,v-i} R^{(1)}_{n,i}\\
     & +\sum_{m\neq p}  \sum_i k_{v-i}(k_i+k_{v-i})f_{p-m,v-i} R^{(1)}_{m,i}  ]\nonumber-\beta^2 \bar{f}_0^2 k_v^2 R^{(2)}_{p,v}
\end{split}
\label{eq:4-ord-2-fourier}
\end{equation}
whose solution, using $k_i+k_{v-i}=k_v$ and dividing by $k_v^4$, reads
\begin{equation}
\begin{split}     
& R^{(2)}_{p,v}=\\
&-\dfrac{\mathlarger{\sum}\limits_{n\neq p} \mathlarger{\sum}\limits_{i}  R^{(1)}_{n,i} \dfrac{\beta f_{p-n,v-i}}{k_v}\left[2\iota -\iota \dfrac{k_{v-i}}{k_v}\dfrac{k_{v-i}-2k_i}{k_v}-3\left(\dfrac{\beta \bar{f}_0}{k_v}\right)^2\right]}{\Gamma_{p,v}+2\iota \dfrac{\beta \bar{f}_0}{k_v}}
\end{split}
    \label{eq:4-ord-2-fourier-simpl}
\end{equation}
where $\Gamma_{p,v}$ has been defined in Eq.~\eqref{eq:app-Gamma}.

Finally, we compute the flux at different orders of the expansion. 
\begin{enumerate}
\item At zeroth order Eq.~\eqref{eq:J-app} reads:
\begin{align}
    J^{(0)}=\beta D \bar{\rho}\bar{f}_0
\end{align}

\item At first order, Eq.~\eqref{eq:J-app} is:
\begin{equation}
\begin{split}     
J^{(1)}= & 2\frac{\beta D}{L}\sum_{p}\sum_v \bar{\rho}\beta ^2 (\text{Re}(f_{p,v})^2+ \text{Im}(f_{p,v})^2).\\
& (\frac{\beta \bar{f}_0}{k_v^2}\frac{\Gamma_{p,v}-1}{\Gamma_{p,v}^2+4\left(\frac{\beta f_0}{k_v}\right)^2})
\end{split}
\end{equation}
Introducing the amplitude of the mode $f_{p,v}$ 
\begin{align}
 |f_{p,v}|^2\equiv \text{Re}(f_{p,v})^2+\text{Im}(f_{p,v})^2
\end{align}
the last expression reads:
\begin{align}
    J^{(1)}= 2\frac{\beta D}{L}\sum_{p}\sum_v \bar{\rho}\beta ^2 |f_{p,v}|^2 \frac{\beta \bar{f}_0}{k_v^2}\frac{\Gamma_{p,v}-1}{\Gamma_{p,v}^2+4\left(\frac{\beta f_0}{k_v}\right)^2}
    \label{eq:J1-app}
\end{align}
which corresponds to  Eq.3 of the main text.
\item At second order Eq.~\ref{eq:J-app} reads:
\begin{align}
    J^{(2)}= 2\frac{\beta D}{L}\sum_p \sum_v \text{Re}(R^{(2)}_{p,v}f^*_{p,v})
\end{align} 
which can be identified with Eq.5 of the main text.
\end{enumerate}

\newpage
\bibliography{main}

\end{document}